\documentclass[journal=None,manuscript=article]{achemso}

\usepackage[version=3]{mhchem} % Formula subscripts using \ce{}
\usepackage{amsmath}
\usepackage[normalem]{ulem}
\usepackage{amssymb}
\usepackage{subfigure}
\usepackage{graphicx}
\usepackage{changes}

\author{Yixuan Zhang}
\affiliation{Institut of Materials Science, Technical University of Darmstadt, 64287, Darmstadt, Germany}

\author{Teng Long}
\affiliation{School of Materials Science and Engineering, Shandong University, 250061, Jinan, China}

\author{Hongbin Zhang}
\affiliation{Institut of Materials Science, Technical University of Darmstadt, 64287, Darmstadt, Germany}
\email{hongbin.zhang@tu-darmstadt.de}

\title[\textsf{achemso} SDMS]
 {Stable diffusion for the inverse design of microstructures}
\abbreviations{Microsturcture, diffusion, generative deep learning}

\begin{document}

%%%%%%%%%%%%%%%%%%%%%%%%%%%%%%%%%%%%%%%%%%%%%%%%%%%%%%%%%%%%%%%%%%%%%
%% The "tocentry" environment can be used to create an entry for the
%% graphical table of contents. It is given here as some journals
%% require that it is printed as part of the abstract page. It will
%% be automatically moved as appropriate.
%%%%%%%%%%%%%%%%%%%%%%%%%%%%%%%%%%%%%%%%%%%%%%%%%%%%%%%%%%%%%%%%%%%%%

%%%%%%%%%%%%%%%%%%%%%%%%%%%%%%%%%%%%%%%%%%%%%%%%%%%%%%%%%%%%%%%%%%%%%
%% The abstract environment will automatically gobble the contents
%% if an abstract is not used by the target journal.
%%%%%%%%%%%%%%%%%%%%%%%%%%%%%%%%%%%%%%%%%%%%%%%%%%%%%%%%%%%%%%%%%%%%%
\begin{abstract}
In materials science, microstructures and their associated extrinsic properties are critical for engineering advanced structural and functional materials, yet their robust reconstruction and generation remain significant challenges. In this work, we developed a microstructure generation model based on the Stable Diffusion (SD) model, training it on a dataset of 2D 576,000 synthetic microstructures containing both phase and grain orientation information. This model was applied to a range of tasks, including microstructure reconstruction, interpolation, inpainting, and generation. Experimental results demonstrate that our image-based approach can analyze and generate complex microstructural features with exceptional statistical and morphological fidelity. Additionally, by integrating the ControlNet fine-tuning model, we achieved the inverse design of microstructures based on specific properties. Compared to conventional methods, our approach offers greater accuracy, efficiency, and versatility, showcasing its generative potential in exploring previously uncharted microstructures and paving the way for data-driven development of advanced materials with tailored properties.
\end{abstract}

%%%%%%%%%%%%%%%%%%%%%%%%%%%%%%%%%%%%%%%%%%%%%%%%%%%%%%%%%%%%%%%%%%%%%
%% Start the main part of the manuscript here.
%%%%%%%%%%%%%%%%%%%%%%%%%%%%%%%%%%%%%%%%%%%%%%%%%%%%%%%%%%%%%%%%%%%%%
\section{Introduction}

The development of advanced structural and functional materials is crucial for addressing global challenges such as energy scarcity and the ever-increasing demands of information technologies\cite{agrawal_perspective_2016}. At the heart of materials science lies the intricate composition-processing-structure-property (CPSP) relationships, encompassing composition, processing, crystal-/micro-structure, and material properties\cite{debroy_additive_2018, krauss_steels_2015}. While machine learning has made significant strides in quantifying this chain in the forward way, from identifying synthesis recipes\cite{jain_commentary_2013} to modeling physical properties\cite{schmidt_recent_2019}, a complete quantitative mapping of the CPSP relationships remains challenging. The emergent data-driven approaches based on machine learning focus mostly on make predictions in a forward inference way\cite{butler_machine_2018, ladani_applications_2021, huo_machine-learning_2022}. There have been limited efforts to explore the CPSP relationships in the reversed way, which is essential to accomplish the inverse design of advanced materials\cite{zunger_inverse_2018}. For instance, the proposed "inverse design"\cite{de_pablo_materials_2014, de_pablo_new_2019} aims at shifting the focus from deriving properties from structures to predicting structures with desired properties, including both the crystal structures\cite{long_constrained_2021, long_inverse_2022, yang_scalable_2023, zeni_mattergen_2024} and microstructures\cite{iyer_conditional_2019, chun_deep_2020, lambard_generation_2023, dureth_conditional_2022, lee_microstructure_2023}. Such an inverse design paradigm, leveraging techniques like high-throughput combinatorial screening, global optimization, and generative deep learning\cite{sanchez-lengeling_inverse_2018, noh_inverse_2019}, promises to accelerate material discovery and engineering more efficiently.

In particular, the quest for understanding and manipulating microstructure has long been pursued by both experimental and simulational methods. These methods, while instrumental, primarily focus on identifying key features within microstructure to elucidate material properties and guide the design of new materials. Experimental techniques can be applied to characterize two-dimensional (2D) and three-dimensional (3D) microstructural morphology at varying resolutions, {\it e.g.}, from sub-nanometer using atom probe tomography (APT)\cite{gault_atom_2021} to $\mu$m using scanning electron microscopy (SEM)\cite{hawkes_science_2007, humphreys_electron_2014}. Corresponding scale-bridging simulations, ideally carried out integrating accurate density functional theory (DFT), molecular dynamics (MD), phase field (PF) and finite element modelling, can in principle quantify the underlying mechanisms but have been mostly performed for selected representative microstructures. The bottleneck lies in a reliable microstructure characterization and reconstruction (MCR), usually done based on statistical functions like two-point correlation and linear-path functions and statistical physical descriptors like cluster’s nearest center distance and orientation angle of a cluster’s principle axis\cite{bostanabad_computational_2018}, which are available in established tools, such as Dream3D\cite{groeber_dream3d_2014}, OptiMic\cite{serrao_optimic_2021} and MCRpy\cite{seibert_microstructure_2022}. However, while these descriptors offer valuable insights, they often remain confined to specific microstructure systems, potentially overlooking intricate features or complex interdependencies within the microstructure. For instance, it is demonstrated that segmentation variations propagate into the simulated physical properties where complicated uncertainty quantification is indispensable to elucidate the microstructure-extrinsic property mapping\cite{krygier_quantifying_2021}. 

To tackle such challenges, the advent of computer vision (CV) technologies can herald a transformative shift\cite{chai_deep_2021, xu_computer_2021}, in particular generative deep learning models offer a holistic approach to microstructure design by comprehending the distribution of microstructures in the visual phase space\cite{holm_overview_2020}. 
However, building a robust uniform generative model necessitates a vast and diverse dataset, ideally encompassing over 20,000 samples. While existing microstructure image databases, such as NFFA-EUROPE SEM Dataset and ASM Micrograph Database\cite{NFFA_EUROPE, lupulescu_asm_2015}, offer a plethora of high-quality images, they are still not adequate for the scale required for a comprehensive generative model. Currently, the generative models can be broadly classified into three classes based on their size of dataset, focus and application: the local feature generative models, the regional models and the uniform generative models.

The local feature generative models, typically trained on a limited dataset with less than 10 microstructure images, emphasize pinpointing features in the visual phase space. Leveraging convolutional layers, such models can capture the underlying statistical features to reconstruct or generate microstructures. While the resulting microstructure might deviate from the original, their morphological statistical distribution remains consistent. Such models are predominantly employed for constructing larger microstructures for simulations\cite{chun_deep_2020, gayon_lombardo_pores_2020, hsu_microstructure_2021,henkes_three_dimensional_2022, kench_microlib_2022} or rectifying experimental anomalies\cite{fokina_microstructure_2020, squires_two_2022}. Generative Adversarial Networks (GANs), which operate without an explicit density function, learn through a zero-sum game between the generator and discriminator, are commonly used here, but their lack of an explicit density function can pose challenges both in training and inference.

The regional generative models, mostly used in catering to specific regions in the visual phase space, are trained on datasets ranging from hundreds to thousands of microstructure images across diverse morphologies of a particular compound system. Their strength lies in capturing the local morphological distribution, offering valuable insights into the process-structure-property linkage. Both GANs and variational autoencoders (VAEs) are employed here. For instance, VAEs, designed to learn an explicit density function in the latent space, offer an effective feature representation\cite{cang_improving_2018, kim_exploration_2021}. However, their assumption of data following a multivariate Gaussian distribution can sometimes lead to suboptimal results. While models can be improved by using more complex priors or learnable priors, one needs to carefully weigh the balance between model effectiveness and the increased complexity of datasets, training and sampling\cite{klushyn_learning_2019, tomczak_vae_2018}. In contrast, GANs \cite{yang_microstructural_2018, iyer_conditional_2019, ma_image_driven_2020, lee_virtual_2021, lambard_generation_2023}, while powerful, can sometimes lead to optimization instabilities.

The uniform generative models are designed for a comprehensive understanding of the global data distribution. These models necessitate expansive datasets and stable architectures. In recent years, the evolution of generative models has been marked by endeavors to harness the strengths of both VAEs and GANs. However, the overarching question has been whether a model can be developed that combines the benefits of these two categories? The denoising diffusion probabilistic model (DDPM), as demonstrated in various studies, offers a potential solution\cite{sohl_dickstein_deep_2015, ho_denoising_2020, nichol_improved_2021, dhariwal_diffusion_2022, peebles_scalable_2022, croitoru_diffusion_2022, song_denoising_2022}. The DDPM model highlights the continuous advancements in the realm of microstructure generative models, in works like that of Düreth, C., et al.\cite{dureth_conditional_2022}, DDPM was trained on extensive datasets, showcasing its potential in capturing intricate microstructural morphologies. However, during image generation, DDPM models usually operate directly in pixel space, which leads to high computational complexity during model training and high inference cost due to sequential evaluation, which in turn limits the resolution or size of the generated microstructures to a low range. In addition, pixel-level diffusion, while guaranteed to preserve the spatial detail of an image, may not be reliable in characterizing the broad contextual features. 

The Stable Diffusion (SD), also referred to as the Latent Diffusion Model (LDM), presents distinct advantages over DDPM\cite{rombach_high_resolution_2022}. The emergence of Latent Diffusion Models (LDM) is based on the idea that semantic and perceptual information exist at different image scales. In microstructure images, semantic information refers to the meaningful interpretation of features, phases, and patterns, which relate to material properties, while perceptual information focuses on visual aspects like fidelity and resolution. In a unified microstructure generation model, understanding semantic information is crucial for studying structure-property linkages, alongside maintaining good visual quality. Traditional diffusion models (DDPM) are limited by their computationally intensive pixel-level evaluations, which restrict scalability and semantic diversity. LDM overcomes this by separating semantic and perceptual compression using autoencoders and diffusion in latent space, significantly reducing complexity and computational costs. This enables more efficient processing of intricate microstructures while preserving essential details. Additionally, LDM’s use of cross-attention layers enhances high-resolution synthesis and allows for conditional generation across multiple models. For instance, the property control can be easily introduce into SD model by combining the ControlNet\cite{zhang_adding_2023}. In summary, LDM provides a more efficient and versatile solution than DDPM for generating high-fidelity, high-resolution microstructure images. 

In this study, we investigate the capabilities of the SD process, specifically tailored for microstructure reconstruction and generation. Our proposed methodology seamlessly integrates both phase and grain reconstruction and generation within microstructures. By harnessing the inherent advantages of SD, particularly its efficiency in latent space operations and high-resolution synthesis, we aspire to pioneer advancements in the inverse material design. This approach not only offers a robust mechanism for generating intricate microstructures but also holds promise in elucidating the intricate CPSP relationships, paving the way for a more comprehensive understanding of material behaviors.

\section{Methods}

\subsection{Synthetic Microstructure database preparation}

To meet the substantial requirement on data for the SD model, and to fulfill our objectives of reconstructing and generating microstructure with both phase and grain orientation, we opted to utilize synthetic microstructure. In this way, the capabilities of the SD model as a comprehensive generative tool can be demonstrated, and it enables us to compile extensive datasets encompassing the diverse range of information needed. It is noted that further experimental microstructure can be considered later using our trained model.

The generation of synthetic microstructures was facilitated by using Dream3D\cite{groeber_dream3d_2014}, a versatile open-source software designed for the reconstruction, instantiation, quantification, meshing, management, and visualization of multidimensional data. A total number of 1,800 distinct 3D 2-phase microstructures are generated, each characterized by unique statistical properties including grain type (equiaxed, rolled, recrystallized), crystal symmetry, volume fraction $\in(0, 1)$, log grain size distribution ($\mu\in(0.1,7)$, $\sigma\in(0.1,0.75)$). Consequently, pyvista was employed to create 2D slices\cite{sullivan_pyvista_2019}. For each 3D geometry, 128 slices are obtained along each of the x, y, and z axes, resulting in a comprehensive dataset of 691,200 phase images (containing only bi-phase morphology) and an equal number of grain images (w/ crystalline directions). From this dataset, 576,000 (83.33\%) images were designated as the training set, with the remaining 115,200 images forming the validation set. Such a systematic approach for data generation and slicing ensures a robust foundation for training our diffusion model and hence balanced evaluation of its performance in microstructure reconstruction and generation.

\subsection{Denoising Diffusion Probabilistic Models}

Generative models aim to establish a mapping from latent to data spaces, facilitating the generation of novel, coherent samples by inverting the mapping from latent distributions to data distributions. The Denoising Diffusion Probabilistic Models (DDPM) incorporates both a diffusion (forward) process and a denoising (reverse) process, each represented by Markov chains that employ Gaussian transformation matrices at each timestep. The diffusion process incrementally introduces Gaussian noise into the data following a predetermined variance schedule. And the denoising process, driven by a U-net model learnt from diffusion process, is initiated from a purely random noise distribution and iteratively refines this noise through a series of timesteps to gradually reduce noise and eventually yield a realistic microstructure images. Consequently, the probability diffusion distribution $q(x_t|x_{t-1})$ of a noised microstructure state $x_t$ at timestep $t$, given the previous state $x_{t-1}$, is mathematically defined as:

\begin{equation}
    q(x_t|x_{t-1}) := \mathcal{N}\big(x_t; \sqrt{1-\beta_t}x_{t-1}, \beta_t\mathbf{I}\big),
\end{equation}
where $\beta_t\in (0,1)$ denotes the scheduled variance at the timestep $t$ in the Markov chain. Correspondingly, the diffusion Markov chain conditioned on microstructure data $x_0$ can then be expressed as:

\begin{equation}
    q(x_{1:T}|x_0) := \prod_{t=1}^Tq(x_t|x_{t-1}),
\end{equation}

The denoising process aims to reverse the diffusion from state $x_T$ to data $x_0$, thus a reverse Gaussian transform conditioned on $x_t$ to get $x_{t-1}$ should be established. In the meanwhile, this reverse transform should preserve all the features in the corresponding forward transform. In this regards, the distribution $p_{\theta}(x_{t-1}|x_t)$, to be fitted by learning, can be formulated as:

\begin{equation}
    p_{\theta}(x_{t-1}|x_t) = \mathcal{N}(x_{t-1}; \mu_\theta(x_t,t), \Sigma_{\theta}(x_t, t)),
\end{equation}
where $\mu_\theta(x_t,t)$ and $\Sigma_{\theta}(x_t, t)$ are the mean and variance of the Gaussian transform predicted by the ML model, $\theta$ denotes the hyperparameters of the model. And the Markov chain of the denoising process is then:

\begin{equation}
 p_{\theta}(x_{0:T}):=p(x_T)\prod_{t=1}^Tp_{\theta}(x_{t-1}|x_t)
\end{equation}

The ML model combining $q$ ad $p$ can be interpreted as a VAE and the loss function can be expressed as an variational lower bound $L_{vlb}$:

\begin{equation}
L_{vlb} := \mathbb{E}_q\Big[-log\frac{p_{\theta}(x_{0:T})}{q(x_{1:T}|x_0)}\Big]\ge\mathbb{E}\big[-logp_{\theta}(x_0)\big]
\end{equation}

After straightforward derivations\cite{ho_denoising_2020, nichol_improved_2021}, it is obvious that the essense of $L_{vlb}$ is to let the model to reversely mimic the diffusion at each timestep, \textit{i.e.}, to match $p_{\theta}(x_{t-1}|x_t)$ to the reverse form of diffusion $q(x_{t-1}|x_t, x_0)$. Using the Bayes theorem, it yields

\begin{equation}
q(x_{t-1}|x_t, x_0)=q(x_t|x_{t-1}, x_0)\frac{q(x_{t-1},x_0)}{q(x_t, x_0)}=\mathcal{N}(x_{t-1}; \widetilde{\mu}(x_t,x_0), \widetilde{\beta_t}\textbf{I}),
\end{equation}
where $\widetilde{\mu}(x_t,x_0):=\frac{\sqrt{\overline{\alpha}_t-1}}{1-\overline{\alpha}_t}x_0+\frac{\sqrt{\alpha_t(1-\overline{\alpha}_{t-1})}}{1-\alpha_t}x_t=\frac{1}{\sqrt{\alpha_t}}(x_t-\frac{\beta_t}{\sqrt{1-\overline{\alpha}_t}}\epsilon)$ and $\widetilde{\beta_t}:=\frac{1-\overline{\alpha}_{t-1}}{1-\overline{\alpha}_t}\beta_t$. The mean fitted by model can be calculated as a function of $\epsilon_{\theta}(x_t,t)$

\begin{equation}
    \mu_\theta(x_t,t) = \frac{1}{\sqrt{\alpha_t}}\Big(x_t-\frac{\beta_t}{\sqrt{1-\overline{\alpha}_t}}\epsilon_{\theta}(x_t,t)\Big),
\end{equation}
with $\alpha_t = 1-\beta_t$ and $ \overline{\alpha_t} = \prod_{s=0}^t\alpha_s$. Therefore, the loss function of the diffusion model can be simplified into
\begin{equation}
L_{DM} = \mathbb{E}_{x,\epsilon\sim\mathcal{N}(0,1),t}\bigg[\|\epsilon-\epsilon_{\theta}(x_t,t)\|_2^2\bigg]
\end{equation}

The variance can be either fixed to a constant of $\beta_t$ or $\widetilde{\beta_t}$ as in work of Ho {\it et al.}\cite{ho_denoising_2020} to reduce the training cost when using large number of diffusion timesteps (e.g., typically 1000), as $\beta_t$ and $\widetilde{\beta_t}$ in the equation are corresponding to upper and lower bounds for the reverse process variances, respectively. They can also set to be $\Sigma_{\theta}(x_t, t) = exp(\upsilon_{\theta}(x_t,t)log\beta_t+(1-\upsilon_{\theta}(x_t,t))log\widetilde{\beta_t})$ as in work of Nichol and Dhariwal\cite{nichol_improved_2021}, however, a hybrid loss function $L_{hybrid}:=L_{DM}+\lambda L_{vlb}$ should also be defined as $L_{DM}$ does not contain the variance information.

\subsection{Stable Diffusion}

SD models distinguish between semantic and perceptual information in microstructure images by employing autoencoders and DDPM on latent representations to overcome the computational limitations of traditional DMs.

The autoencoder component in SD consists of an autoencoder that combines perceptual loss and patch-based adversarial target loss. The encoder $\mathcal{E}$ will encode the microstructure data $x$ as the point $z$ on the latent space $z=\mathcal{E}(x)$, and the decoder $\mathcal{D}$ will then try to reconstruct the microstructure from this point $\widetilde{x}=\mathcal{D}(z)=\mathcal{D}(\mathcal{E}(x))$
\begin{equation}
L_{Autoencoder} = \mathop{min}\limits_{\mathcal{E}, \mathcal{D}}\mathop{max}\limits_{\psi}\Big(L_{rec}(x,\mathcal{D}(\mathcal{E}(x)))-L_{adv}(\mathcal{D}(\mathcal{E}(x)))+logD_{\psi}(x)+L_{reg}(x;\mathcal{E},\mathcal{D})\Big),
\end{equation}
where $L_{rec}$ is the reconstruction loss, $L_{adv}$ is the adversaial loss, $L_{reg}$ is the regularizing loss and $D_\psi$ is the discriminator. 

With a low-dimensional latent space generated by a pre-trained autoencoder model that is perceptual equivalent to real microstructural space, it is now possible to train the DM model using the latent representations. 
\begin{equation}
L_{LDM} = \mathbb{E}_{\mathcal{E}(x),\epsilon\sim\mathcal{N}(0,1),t}\bigg[\|\epsilon-\epsilon_{\theta}(z_t,t)\|_2^2\bigg]
\end{equation}
where $\epsilon_{\theta}(z_t,t)$ can be obtained using a time-conditional UNet, with $z_t$ being the latent representation of noised data at timestep $t$.

\subsection{ControlNet}

As microstructural phases and grains share some common information, such as size, shape, spatial distribution and volume fraction, boundary characteristics, connectivity and continuity, \textit{etc.,} it is possible to train a SD model integrating both phase and grain information together while sharing common latent representations. Our proposed methodology involves the amalgamation of colorful grain images, where the grain orientations are encoded in RGB channels, with grayscale phase images, resulting in a 4-channel input. Such a holistic representation captures both the geometric boundaries and compositional details of the microstructure, which not only ensures both the accurate microstructure analysis and the model generalizbility, but also allows the model to better handle a variety of scenarios and downstream tasks related to variety of property predictions. However, introducing a fourth channel to traditional SD models, designed for RGB images, requires architectural adjustments and precise calibration of the noise injection during diffusion stages, to ensure the model can accurately capture the complex relationship between grain orientation and phase information without compromising essential details.

The U-net structure in SD model, which inherently captures grain orientation and phase information, enables the reconstruction of grain orientations using only phase data. However, adapting the SD model's conditioning to effectively use the 4-channel input without compromising its generality requires careful consideration. Directly incorporating conditions via U-net's cross-attention layer introduces additional computational and labeling costs and can reduces the generality of the model by adding additional condition-related weights to the current model and skews other usual weights in the model toward the current condition. To maintain the model's versatility for various conditions without adding undue complexity, ControlNet, a specialized conditioning technique that optimizes the use of combined microstructural information, was employed, which ensures the model's broad applicability and ease of future fine-tuning.

\begin{figure}
\centering
  \includegraphics[width=15cm]{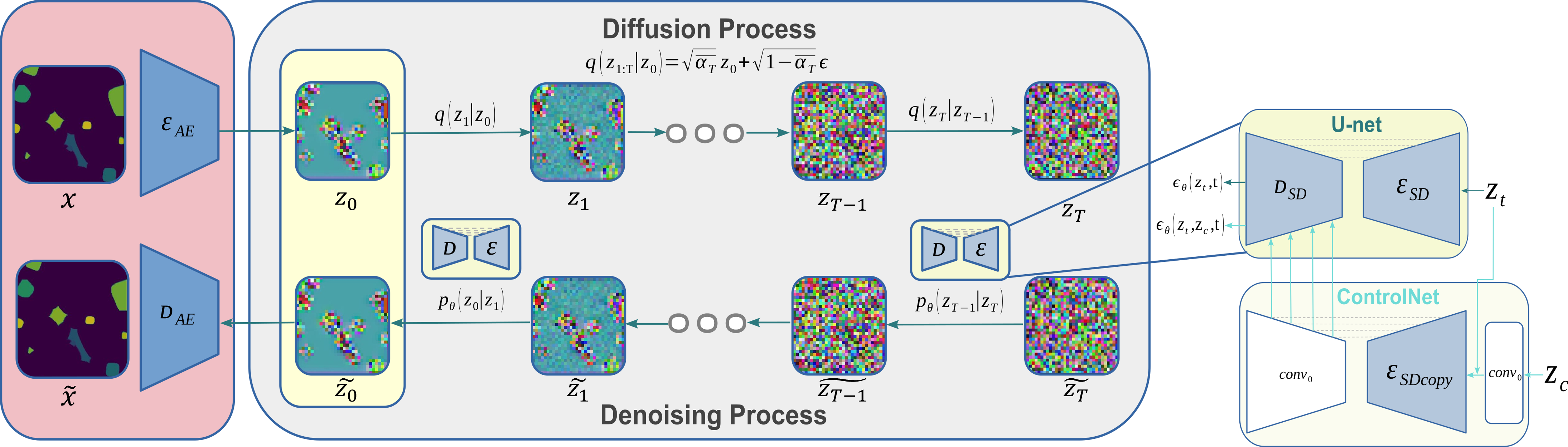}
  \caption{The workflow of our microstructure SD model.}
  \label{fgr:SD_workflow}
\end{figure}

ControlNet acts as a guiding mechanism within the SD framework\cite{zhang_adding_2023}. It influences the generation process by steering the model towards producing outputs that align with specific desired characteristics or conditions. Such additional conditions could be anything from text or image descriptions to specific material properties in the case of microstructure generation. This conditioning allows for more targeted generation. In the diffusion process, ControlNet can contribute to the iterative refinement of the generated results by continuously adjusting and guiding the generation at each timestep of the diffusion process, to ensure that final output closely matches the desired outcome, both in terms of visual quality and adherence to specified conditions. In the SD model, ControlNet is integrated into the U-net blocks in order to introduce additional conditions. It use a trained neural block $F(\cdot; \Theta)$ to transform an input feature map $x$ into another feature map $y$  with parameters $\Theta$, {\it i.e.},
\begin{equation}
     y = F(x; \Theta)
\end{equation}
In the implementation of ControlNet, the original parameters $\Theta$ are frozen, and a trainable copy of the block with parameters $\Theta_c$ is created, which accepts an external conditioning vector $c$. ControlNet employs zero convolution layers $Z(\cdot; \cdot)$, which are $1 \times 1$ convolutions initialized with zero weights and biases. The conditioned output $y_c$ is computed as:
\begin{equation}
     y_c = F(x; \Theta) + Z(F(x + Z(c; \Theta_{z1}); \Theta_c); \Theta_{z2})
\end{equation}
where $\Theta_{z1}$ and $\Theta_{z2}$ are the parameters of zero convolution layers before and behind trainable copy block, respectively. The LDM model with ControlNet can be expressed as:
\begin{equation}
L_{LDM}:=\mathbb{E}_{\mathcal{E}(x),z_c,\epsilon\sim\mathcal{N}(0,1),t}\Big[\|\epsilon-\epsilon_{\theta}(z_t,z_c,t)\|_2^2\Big]
\end{equation}
where $z_c = \gamma_{c}(c)$ is the latent condition transferred from input condition $c$ by a pre-trained model $\gamma_{c}$. The overall workflow of our SD model is shown in \ref{fgr:SD_workflow}

\section{Results and discussion}

\subsection{Microstructure reconstruction}

In the SD model, the division of image understanding and reconstruction into semantic and perceptual components plays a crucial role, particularly in the context of microstructures. The perceptual component is essential for capturing the local details of microstructures, making the accuracy of the autoencoder critical. The architecture of the autoencoder applied is specifically adjusted to handle the complex patterns present in our diverse microstructure dataset, which includes a variety of morphologies in two-phase polycrystalline structures.

To quantitatively assess the performance of our autoencoder in reconstructing microstructure images, we evaluated both model metrics and the quality of the reconstructed images. Key metrics such as Root Mean Squared Error (RMSE), Peak Signal-to-Noise Ratio (PSNR), Structural Similarity Index (SSIM), and Feature Similarity Index (FSIM) were employed. These metrics provided a comprehensive view of the model's reconstruction capabilities, capturing pixel-level differences as well as texture changes related to visual perception. This multi-faceted approach offers valuable insights into the model's ability to preserve subtle microstructural details that may not be fully reflected by individual quantitative metrics.

\begin{figure}
\centering
  \includegraphics[width=15cm]{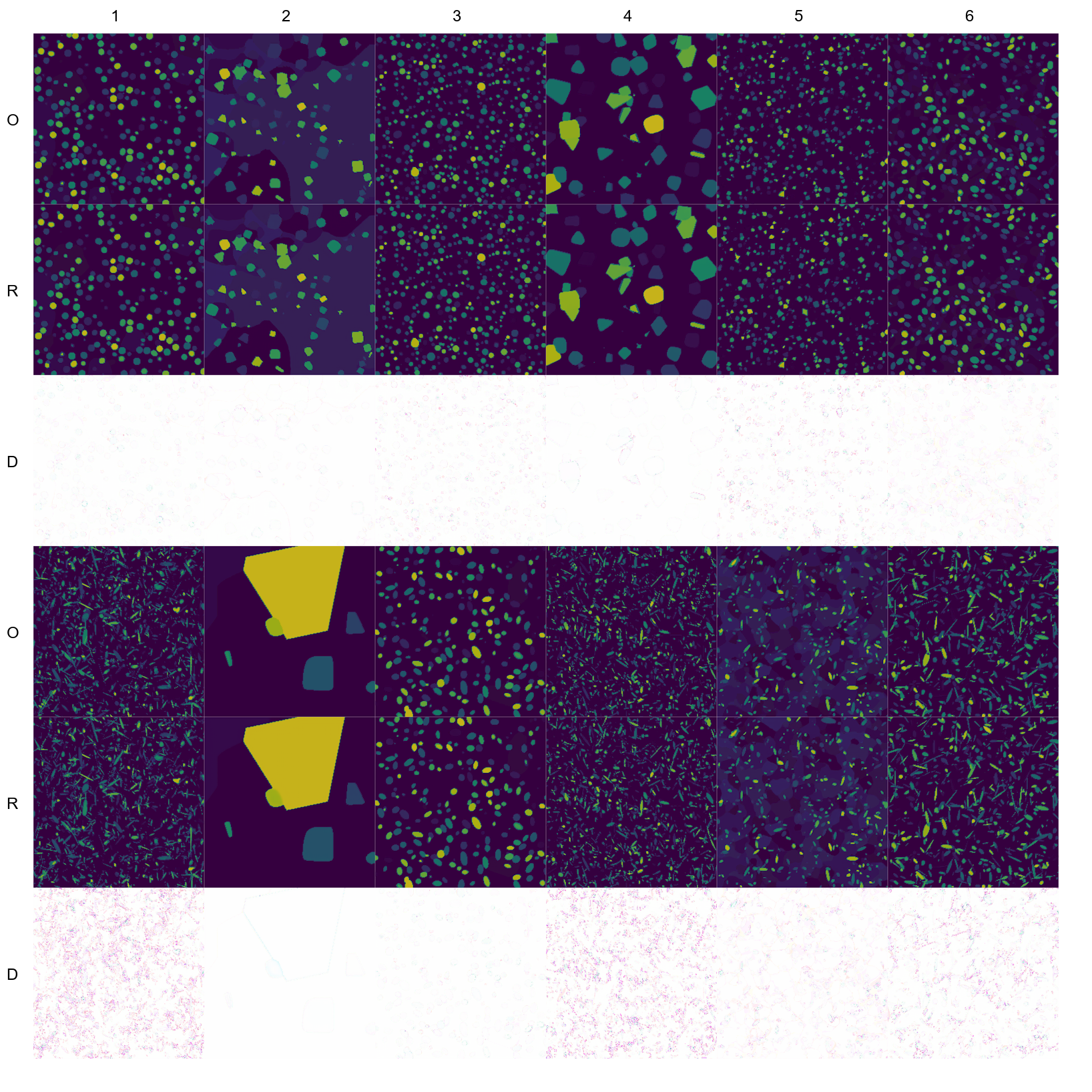}
  \caption{The reconstructed grain image with 1,4 lines the original image, 2,5 lines the reconstructed images and 3,6 lines the error map}
  \label{fgr:euler_reconstruction}
\end{figure}

\begin{figure}
\centering
  \includegraphics[width=15cm]{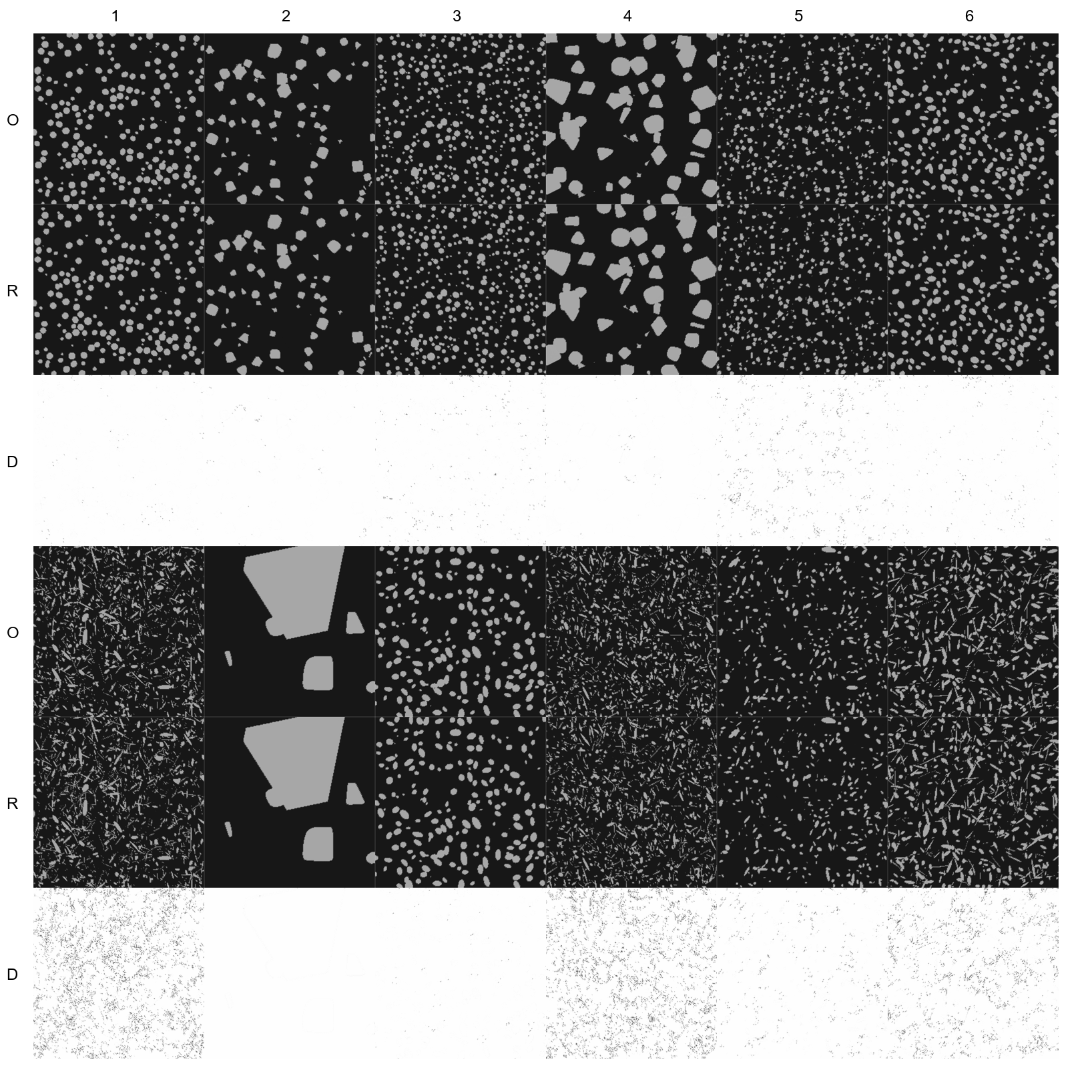}
  \caption{The reconstructed phase image with 1,4 lines the original image, 2,5 lines the reconstructed images and 3,6 lines the error map}
  \label{fgr:phase_reconstruction}
\end{figure}

As shown in Fig. \ref{fgr:euler_reconstruction} and Fig. \ref{fgr:phase_reconstruction}, the visual differences between the original (row 1, 4) and reconstructed images (row 2, 5) are indistinguishable. Only in the error maps (lines 3, 6), which show the absolute differences between pixels, non-blank scatters can be found, which are mostly located at the boundaries of grains or phases in the microstructure, in particular when the grain sizes are smaller than 1 $nm$. It is noted that the spatial resolution of our microstructure images is 0.5 nm, and the typical mesh grid for micromagnetic simulations is 1 nm. Therefore, it is suspected that the deviations in the sub-nm regime will lead to marginal influence on the phyical properties. Nevertheless, we would like to emphasize that the spatial resolution in the SD model can be further increased, depending on the corresponding resolution in the training datasets.

In Table \ref{tbl:reconstruction_metric}, a quantitative assessment of the reconstructed image quality is presented, utilizing 27,696 (24\%) randomly selected image pairs from the validation set. Among four metrics considered, RMSE quantifies the average magnitude of errors between the original and reconstructed images, where lower values signify higher similarity. PSNR, representing the ratio between the maximum possible signal power and the power of corrupting noise, indicates better quality with higher values. SSIM is more aligned with human visual perception as it considers changes in texture, luminance, and contrastranging. It ranges from 0 to 1, with 1 denoting perfect similarity. FSIM, like SSIM, is also used to assess the image similarity, targeting more on the extraction and comparison of significant features like edges, corners, and other key structural elements in the images. It also ranges from 0 to 1, where 1 signifies exact resemblance. 

\begin{table}[h!]
\small
  \caption{\ The average RMSE, PSNR ($dB$), SSIM and FSIM between original and reconstructed images of grain and phase.}
  \label{tbl:reconstruction_metric}
  \begin{tabular*}{\textwidth}{@{\extracolsep{\fill}}lcccc}
    \hline
    Type & RMSE & PSNR ($dB$) & SSIM & FSIM   \\     
    \hline
    Grain images & 0.010 & 40.199 & 0.995 & 0.810 \\
    Phase images  & 0.018 & 35.221 & 0.988 & 0.863 \\
    \hline
  \end{tabular*}
\end{table}

The performance metrics unequivocally affirm the exceptional reconstruction proficiency of our trained autoencoder model. Notably, the RMSE and SSIM metrics approach their theoretical bests, while the PSNR values exceed 35, underscoring a remarkable pixel-level fidelity between the reconstructed and original images. Furthermore, an FSIM score surpassing 0.8 reflects a substantial feature-level resemblance, reinforcing the model's accuracy in capturing essential image characteristics. Therefore, the autoencoder component of our SD model excels in both accuracy and detail preservation. Its performance, validated through a combination of quantitative and qualitative assessments, underscores its reliability and efficiency in extracting perceptual information from images, which is crucial for the subsequent diffusion models.

\subsection{Interpolation between the microstructures}

As indicated by the reliable construction of microstructure, our SD model works based on a latent space with semantic features while reproducing the original microstructure data distribution. Correspondingly, it enables us to obtain a series of images interpolate between two distinct grain images with drastic contrast in the geometry. For our SD diffusion model, the interpolation hinges on the assignment of weights between different microstructures and the integral role played by diffusion timesteps. These weights control transition between the microstructures, while the diffusion timesteps control the depth of detail merging during interpolation. The interpolation can be mathmatical expressed as:

\begin{equation}
\widetilde{x}=\mathcal{D}(p(z_0)\cdot\prod_{t=1}^Tq(z_t|z_{t-1})\cdot\prod_{t=1}^Tp_{\theta}(z_{t-1}|z_t))=\mathcal{D}((\sqrt{\overline{\alpha}_T}z_0+\sqrt{1-\overline{\alpha}_T}\epsilon)\cdot\prod_{t=1}^Tp_{\theta}(z_{t-1}|z_t))
\end{equation}

where $z_0 = \gamma\mathcal{E}(x_1) + (1-\gamma)\mathcal{E}(x_2)$ with $\gamma$ the weights and $T$ the timestep, $x_1$ and $x_2$ are two distinct microstructures. 

The interpolation process involves blending two distinct grain images, each representing a unique microstructure with diversed morphology. The images are first encoded as the latent representations. The weights are assigned to the latent representations, which vary linearly from 0 to 1, to control the contribution of each image in the interpolated output. In this way, it is able to generate a series of intermediate images that gradually transform from one microstructure to another, particularly as shown by the row of images marked as ``Step 0" in Fig. 3. Additionally, the role of diffusion timestep is also explored, which is not only critical in determining the granularity and smoothness of the transitions, but also has a significant impact on the diversity of the interpolated results and the degree of preservation of the original morphology. 

\begin{figure}
\centering
  \includegraphics[width=15cm]{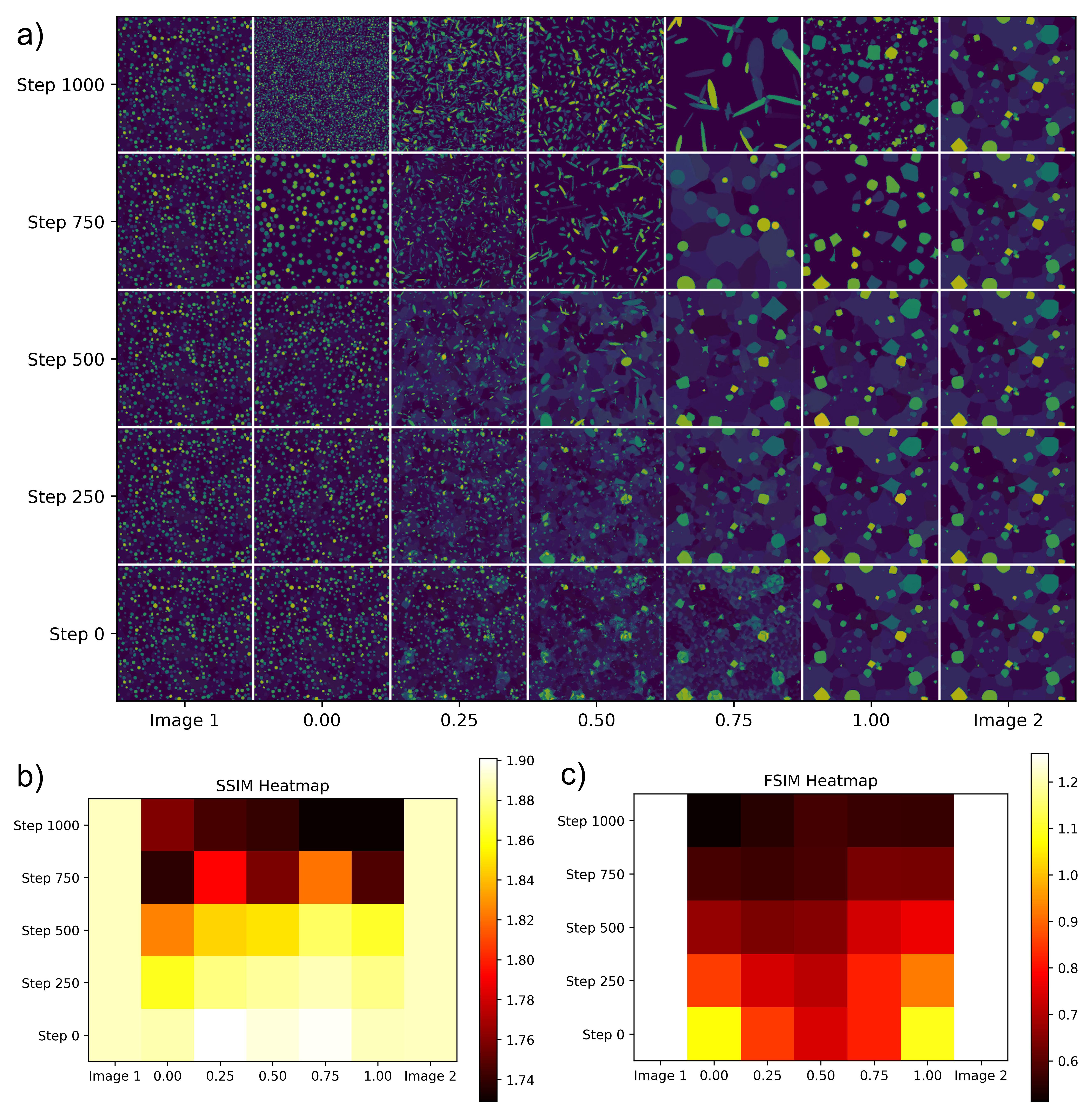}
   \caption{The interpolation of two grain images: a) the visual interpolation with the horizontal axis are the weight between two images ranging from 0 to 1 and the vertical axis are the diffusion timesteps ranging from 0 to 1000; b) the sum of SSIM score and c) the sum of FSIM score of each each interpolation image compared with the two initial grain images. }
\label{fgr:euler_interpolation_metric}
\end{figure}

\begin{figure}
\centering
  \includegraphics[width=15cm]{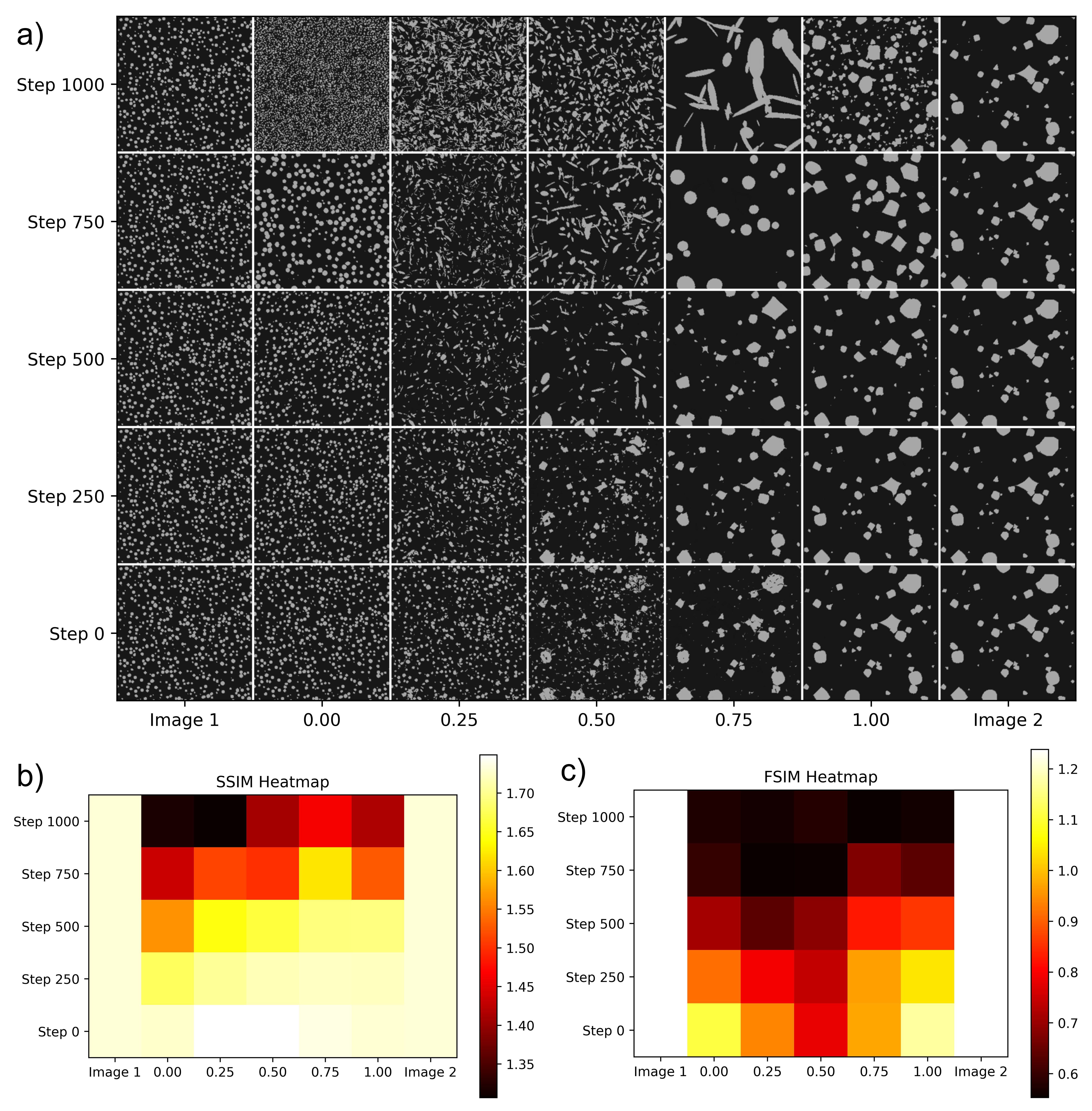}
  \caption{The interpolation of two phase images: a) the visual interpolation with the horizontal axis are the weight between two images ranging from 0 to 1 and the vertical axis are the diffusion timesteps ranging from 0 to 1000; b) the sum of SSIM score and c) the sum of FSIM score of each each interpolation image compared with the two initial phase images. }
\label{fgr:phase_interpolation_metric}
\end{figure}

To assess the quality of our interpolated images, two metrics SSIM and FSIM in the transition are utilized, which are helpful to evaluate the smoothness of the transitions and the fidelity with which the key microstructural features are preserved. High SSIM values indicate that the interpolated images maintain visual similarity, \textit{i.e.}, good structural consistency to the original structures. Whereas high FSIM suggests a close similarity in the distribution of features between the interpolated and original images. In \ref{fgr:euler_interpolation_metric}.a) and \ref{fgr:phase_interpolation_metric}.a), the SSIM values are presented for the grain and phase visual interpolations between two images with different grain sizes and morphology, respectively. The SSIM and FSIM scores for each interpolated image, when compared to the two original images in both grain and phase cases, are summed and presented in \ref{fgr:euler_interpolation_metric}.b) and \ref{fgr:euler_interpolation_metric}.c) and \ref{fgr:phase_interpolation_metric}.b) and \ref{fgr:phase_interpolation_metric}.c), respectively.

From \ref{fgr:euler_interpolation_metric}.a) and \ref{fgr:phase_interpolation_metric}. a), the observed smooth transitions indicate that our model indeed learns a continuously smooth feature space, resulting in realistic and continuous microstructure transitions. When the diffusion time step is small, interpolation between the two images occurs at the pixel level. However, due to the advantages of the autoencoder, these interpolations are not simply a stacking of pixel information, but rather the growth of grains with appropriate boundary shifts during the transition and the consumption of surrounding fine grains. At diffusion time steps above 500, the transition of grains tends to be more on the morphological level, with a gradual smooth transition from 0D fine grains to 1D fibrous grains, and eventually to 2D massive grains. Interestingly, at higher diffusion time steps, the geometrical changes of the grains become more pronounced, while the size changes appear more random. This suggests that the model tends to operate at a semantic level rather than a purely perceptual level at these higher time steps. Such features can also be quantitatively demonstrated in the SSIM and FSIM heatmaps. For instance, according to the SSIM heatmap (Fig. 3b), there exists a roughly "U" shaped color change, which indicates that the larger the diffusion timesteps, the lower the visual similarity. In the FSIM heatmap (Fig. 3c), this "U" shaped distribution is even more pronounced. Since FSIM focuses more on the feature-level comparisons, such a continuous variation with respect to the diffusion timesteps from low to high and in ratios from 0 to 1 does confirms a smooth transition in the underlying space in microstructure SD models.

From the microstructure point of view, the interpolation process offers fascinating insights into potential microstructural transformations or phase transitions. This is particularly relevant in materials science because interpolated images provide us with an potential way to interpret microstructural transformations at the level of model extracted features space. Although this interpolation paths may not correctly represent the gradual changes that may occur in the microstructure during material processing, it shows that the model does provide us with a continuous, smooth microstructural feature space, which ensures that it has the ability to inversely design the microstructural morphology to obtain the desired physical properties, or even processing. This will be saved for future investigation.

\subsection{Microstructure inpainting}

To further demonstrate the capability of our SD model in reconstructing microstructural features, the microstructure images inpaintings are examined for both grain and phase images. This is done by overlaying a central mask, which is half the dimension of the original image in both height and width, onto each image. Consequently, the trained SD model is employed to generate microstructure within the marked region that are compatible with the surrounding original area, leveraging its understanding of the local morphology. That is, during each denoising timestep, the model processes the original region using the diffusion process to introduce noise, while simultaneously denoising the masked region. These two parts are then seamlessly combined to form a complete image for that timestep.

\begin{figure}
\centering
  \includegraphics[width=15cm]{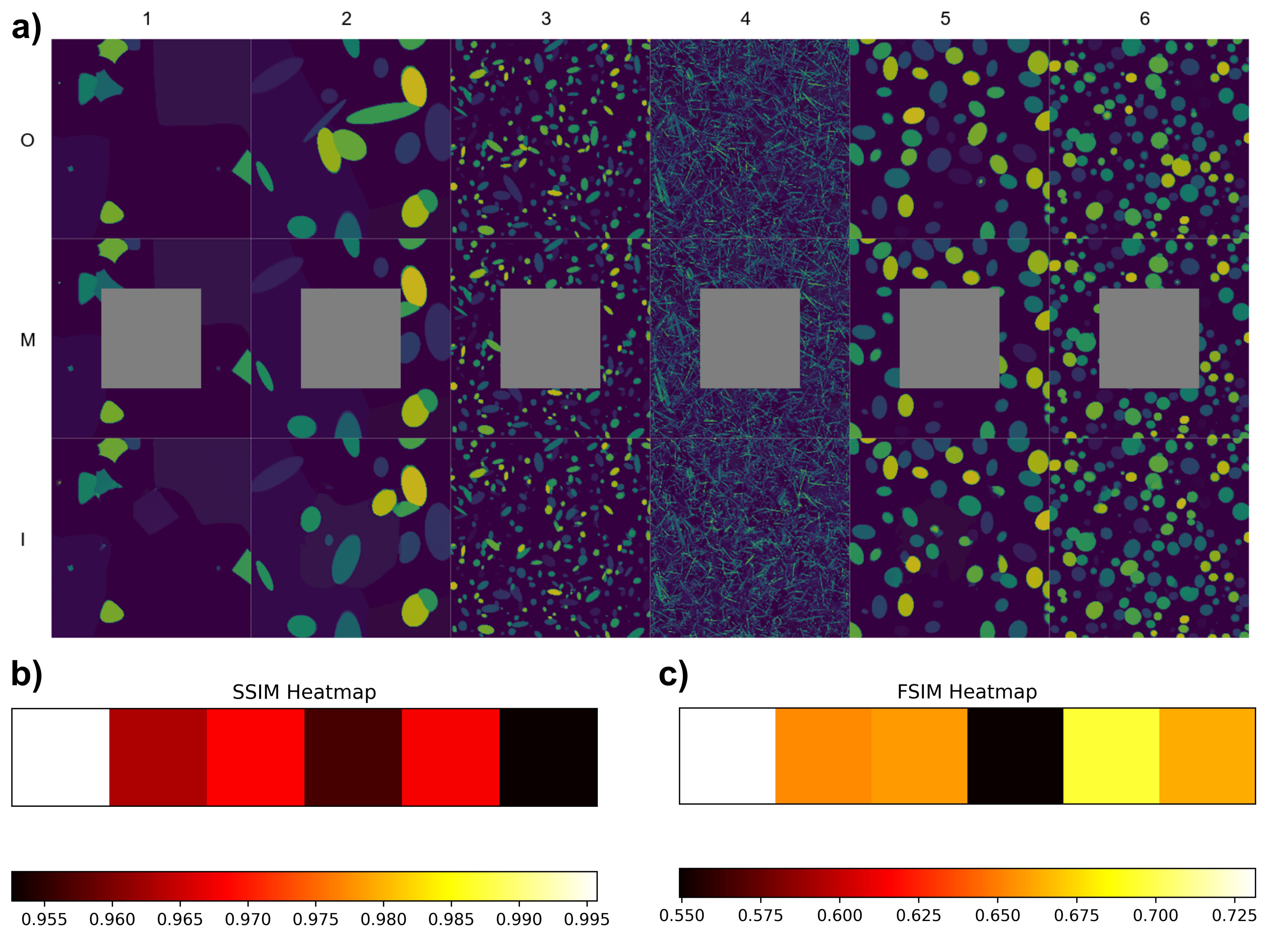}
  \caption{The inpainting of masked grain images: a) the inpainting results with  'O' denotes the original grain images, 'M' denotes masked images, and 'I' represents the inpainted grain images; b) the SSIM score and c) the FSIM score between original images and the inpainting images.}
\label{fgr:euler_inpainting_metric}
\end{figure}

The inpainting results are shown in Fig. \ref{fgr:euler_inpainting_metric}.a, with the corresponding SSIM and FSIM scores in Fig. \ref{fgr:euler_inpainting_metric}.b and c, respectively. Clearly, the transitions between the masked and surrounding regions in the microstructure images are very smooth. In particular, the SD model not only generates morphologies in the masked region that closely match those in the original images, but also generates new grains that match the original morphology and grain orientation of the residual grains, which are located on the boundary and cut by the masked region. In the process of visual comparison, Figures 1,2,5 can find the differences more quickly by looking at the overall changes in phase shape and grain orientation, which corresponds to a more obvious change in perception at the global attention level from the ML perspective. The changes in Figures 3,4,6 require more detailed observation of the global phase density combined with the local grain orientation changes to be noticed. It can be explained that as the grain size decreases and the density becomes larger, the complexity of global attentional perception in the image rises, and therefore more semantic information is needed such as the density and the local detailed perception to distinguish the differences. Quantitatively, the SSIM analysis demonstrated that microstructures 4 and 6 were slightly less similar in terms of texture, brightness, and contrast-range, but still achieved similarity scores of 0.95 or higher. Here it is believed that the variation in luminance and contrast ranging due to small fluctuations in the distribution of phase volume ratios and grain orientations during the generation process, resulting in slightly lower SSIM scores. Similarly, the results of the FSIM analysis also gave a relative lower similarity assessment for microstructure 4, but the similarity scores for the remaining microstructures were all above 0.65. This is because FSIM focuses more on the similarity of shape and structural information of the image. When microstructure 4 is encountered, the model treats needle-like grains with widths smaller than the autoencoder criterion as negligible perceptual information and ignores these small grains during regeneration, resulting in a relative lower FSIM score for this structure.

\begin{figure}
\centering
  \includegraphics[width=15cm]{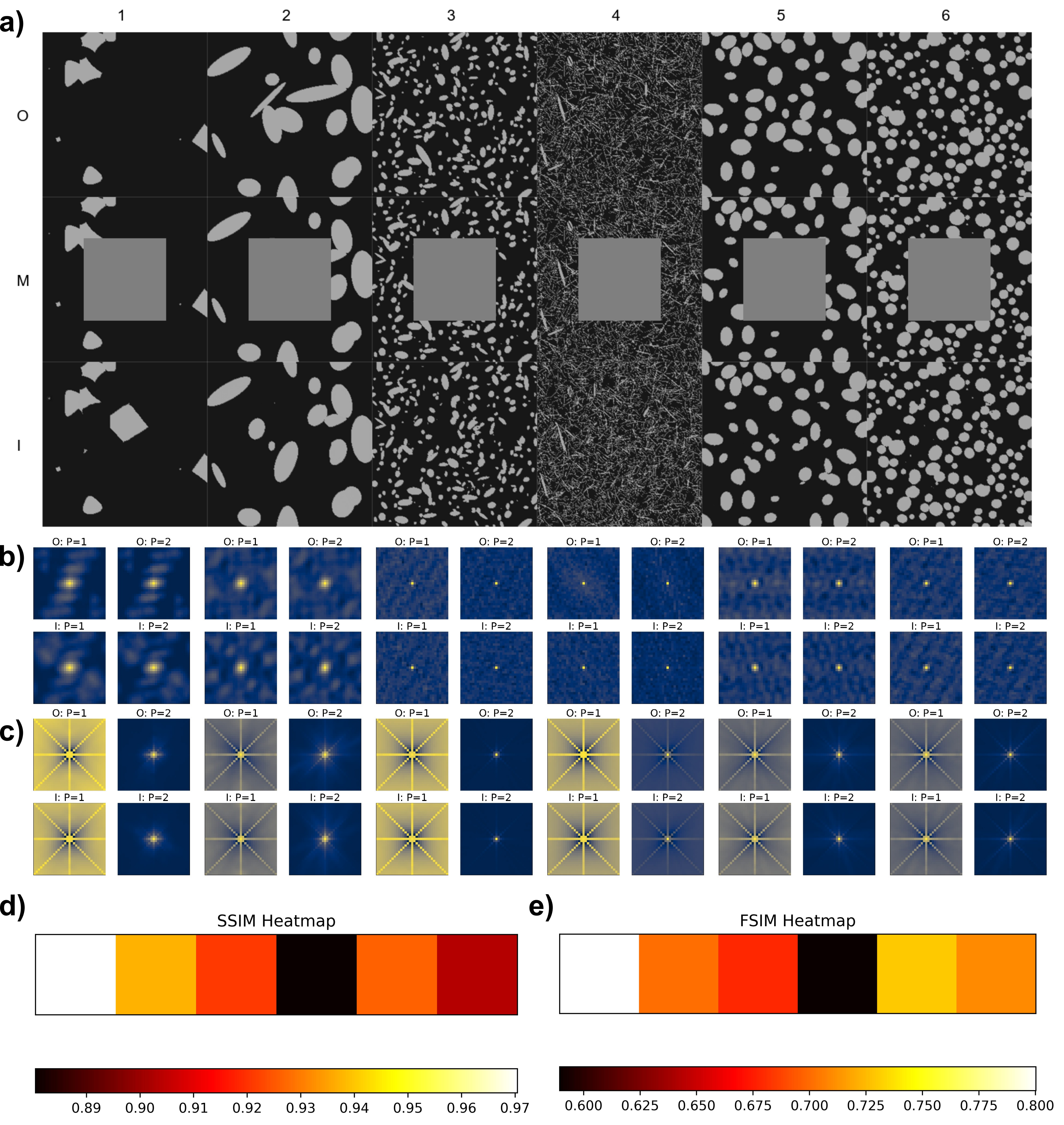}
  \caption{The inpainting of masked phase images: a) the inpainting results with  'O' denotes the original phase images, 'M' denotes masked images, and 'I' represents the inpainted phase images; b) the Two-point correlation plots and c) the Linealpath plots of the original images (O) and the inpainting images (I) of both phases (P), where P=1 being the one with a larger volume fraction, depicted as the darker region in a), and P=2 as the lighter region in a);  d) the SSIM score and e) the FSIM score between original images and the inpainting images.}
\label{fgr:phase_inpainting_metric}
\end{figure}

For the phase inpainting comparison as shown in Fig. \ref{fgr:phase_inpainting_metric}.a), better agreement than the grain images can be obtained, due to the removal of the grain orientation information. The statistical analysis confirms also the visual consistency, as indicated by the FSIM evaluation (cf Fig. \ref{fgr:phase_inpainting_metric}.d). Although the SSIM scores (cf Fig. \ref{fgr:phase_inpainting_metric}.e) is decreased, they are still sufficiently high, marking a high degree of similarity between the original and regenerated morphology. The decrease can be attributed to the fact that as the color information (\textit{i.e.,} the grain orientation information) is lost, the texture information, which can be considered as structural information, becomes more important, leading to a decrease in the SSIM scores.

Visual and quantitative analyses of similarity measurements for both grain and phase microstructure comparison may appear to be contradictive, with less similarity instead requiring longer visual observations to find differences, but in reality they are two sides of the same coin in terms of the effect of microscopic material morphology on properties. In the process of visual characterization, the semantic consistency given by the structure as a whole is more important. That is, the overall distribution of the grains, such as the grain size distribution, the volume fraction, and the phase distribution, have a significant impact on the material properties (including elasticity, plasticity, and thermal conductivity, among others). However, for quantitative analysis, SSIM and FSIM focus more on the perceived differences in local details, such as local grain shape, orientation, and grain boundary behavior of the microstructural morphology.

In addition, the similarity between the original and inpainted phase images using an array of physical descriptors was further evaluated, as facilitated by the MCRpy package\cite{seibert_microstructure_2022}. This analysis encompassed spatial correlation descriptors such as two-point correlations and lineal path correlations, alongside typical descriptors like volume fractions which focuses on volume of different phases and normalized variation which focuses on the local noise level, as well as ML-based descriptors like Gram Matrices. The two-point correlation function quantifies the probability of finding a pair of points at a specific distance apart within the same phase or feature, providing insights into the spatial distribution and homogeneity of microstructures, while the lineal path correlation measures the distribution of uninterrupted line segments within a particular phase, offering valuable information about the continuity and connectivity of microstructural elements. Figure \ref{fgr:phase_inpainting_metric}.b) presents the results of the two-point correlation comparison, while Figure \ref{fgr:phase_inpainting_metric}.c) illustrates the lineal path comparison outcomes. The high similarity between the top and bottom rows of both descriptor maps proves that our inpainted image has high statistical consistency compared with the original image. In the map of two-point correlation, it can be observed that the visual distinguishability of the map distinction decreases with increasing phase distribution homogeneity. The correlation maps in Figures 1 and 2 can pinpoint the different areas relatively quickly with visual comparisons, while Figures 4 and 5 require careful comparisons to find the differences, and Figures 3 and 6 require more focused observations to find the discrepancies. In the lineal path correlation characterization, although the difference is not as pronounced as in the two-point correlation, a similar trend can still be obtained by analyzing the changes in the brightness and contrast of the initial and inpainted images. The analysis with the help of descriptor characterization divided the sample discrimination difficulty into three more detailed categories than the previous visual-only analysis. This is due to the fact that descriptors provide a more accurate characterization of the phase density, volume fraction and distributional homogeneity of the microstructure than direct observation, in other words, descriptors have the ability to better characterize the semantic information of the microstructure. Furthermore, Table \ref{tbl:descriptor_differences} details the mean absolute differences between the original and the corresponding inpainted examples from Figure \ref{fgr:phase_inpainting_metric}.a) for each descriptor. Interestingly, the data from Gram matrices show similar results to our analysis in the previous paragraph, with Figure 3, 6 having the highest similarity, 4, 5 the next highest, and 1, 2 the worst. But on all other descriptors, the data gave the opposite ordering of 1, 2 and 4, 5, even though both two point correlation and linear correlation were consistent with Gram metrics on previous visual similarity determinations of descriptor maps. The consistency between gram metrics and previous analyses can be explained by the fact that the ML descriptor, like us, focuses more on feature similarity between structures, while the rest of the descriptors are more or less influenced by local details. 

In conclusion, the results demonstrate that our diffusion model excels in the inpainting task. Unlike traditional VAE- or GAN-based generative models\cite{lugmayr_repaint_2022}, the diffusion model achieves high-quality inpainting without requiring additional training. This success can be attributed to two key factors: first, the model effectively captures the morphological features of the microstructures; second, the diffusion process, driven by Markov chains, inherently facilitates boundary matching. In this process, the noisy image from the previous timestep serves as a natural boundary condition, ensuring both consistency and completeness in the reconstruction.

\begin{table}[h!]
\small
  \caption{\ The mean absolute differences between the initial images and the inpainted example images for each descriptor, with positive and negative signs in the "volume fraction" indicating an increase or decrease at each phase.}
  \label{tbl:descriptor_differences}
  \begin{tabular*}{\textwidth}{@{\extracolsep{\fill}}lcccccc}
    \hline
    Type & TwoPoint & Lineal & Volume & Variation & Gram  \\
     & Correlations & Path & Fractions &  & Matrices  \\  
    \hline
    Img1-phase1 & 0.0494 & 0.0470 & $-$0.0294 & 0.0134 & 0.0672 \\
    Img1-phase2 & 0.0106 & 0.0090 & $+$0.0294 & 0.0134 & 0.0558 \\
    Img2-phase1 & 0.0327 & 0.0338 & $+$0.0223 & 0.0093 & 0.0675 \\
    Img2-phase2 & 0.0131 & 0.0100 & $-$0.0223 & 0.0093 & 0.0559 \\
    Img3-phase1 & 0.0049 & 0.0064 & $+$0.0034 & 0.0028 & 0.0206 \\
    Img3-phase2 & 0.0017 & 0.0008 & $-$0.0034 & 0.0028 & 0.0162 \\
    Img4-phase1 & 0.0554 & 0.0638 & $-$0.0300 & 0.0208 & 0.0270 \\
    Img4-phase2 & 0.0057 & 0.0038 & $+$0.0300 & 0.0208 & 0.0268 \\
    Img5-phase1 & 0.0466 & 0.0533 & $+$0.0316 & 0.0133 & 0.0446 \\
    Img5-phase2 & 0.0157 & 0.0105 & $-$0.0316 & 0.0133 & 0.0393 \\
    Img6-phase1 & 0.0040 & 0.0034 & $+$0.0013 & 0.0048 & 0.0286 \\
    Img6-phase2 & 0.0031 & 0.0016 & $-$0.0013 & 0.0048 & 0.0218 \\
    \hline
  \end{tabular*}
\end{table}

\subsection{Microstructure generation}

\begin{figure*}[htp]
 \centering
 \includegraphics[width=16cm]{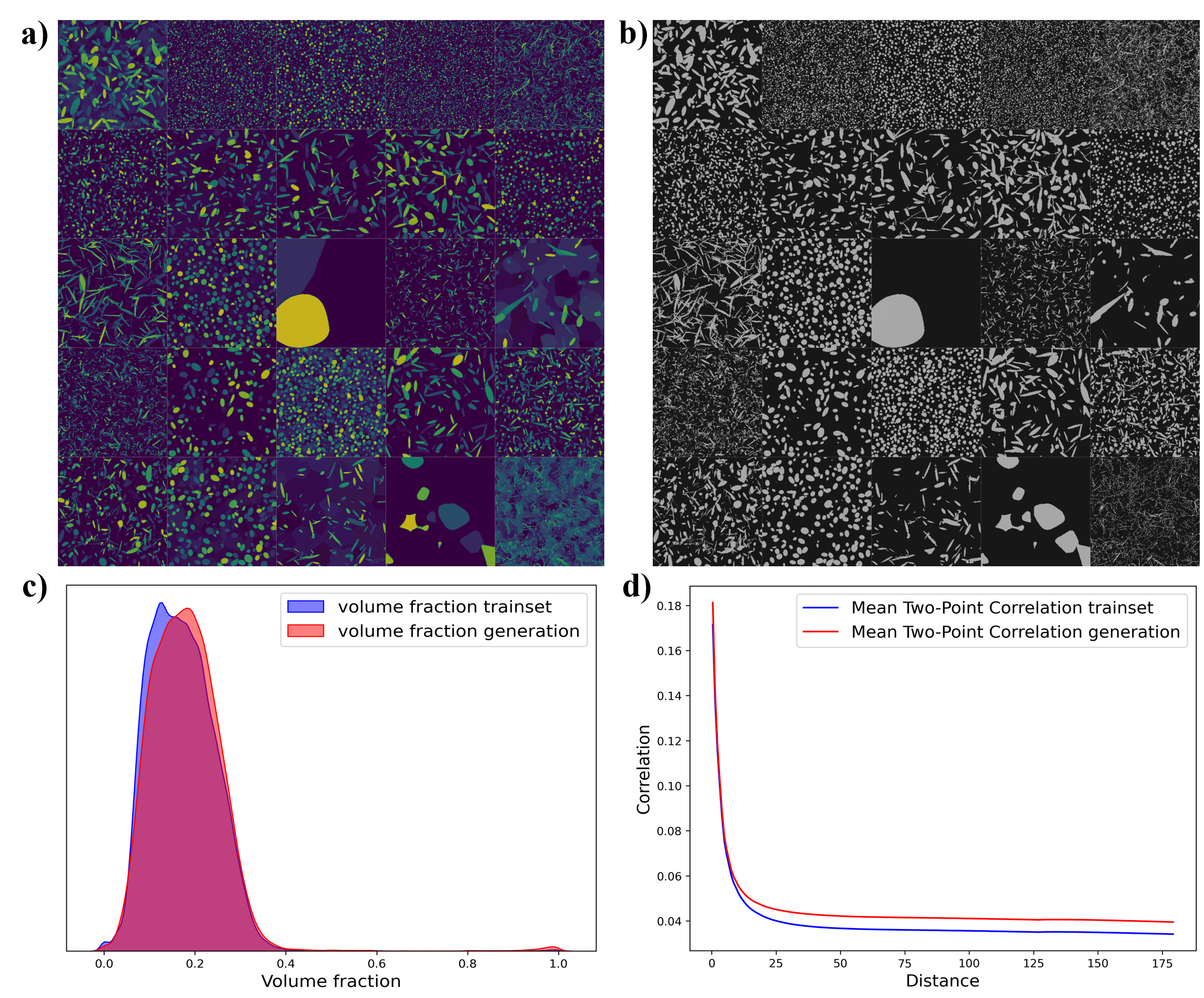}
 \caption{The random generated a) grain images; b) phase images; and the statistical comparison between trainset and generated microstructures of c) the average two-point correlations; d) distribution of volume fraction.}
 \label{fgr:sample_euler_phase}
\end{figure*}

Turning now to the microstructure generation, as illustrated in Figure \ref{fgr:sample_euler_phase}, our SD model successfully generates a variety of microstructures, including fine, fibrous, and bulk grains. Notably, these generated structures are not mere replications of the existing entries within our database, rather they represent distinct instances that expand the current repertoire of the known microstructure morphologies in database. A particularly noteworthy evidence of our model's capability is the microstructure image located at in the third row and fifth column in Figure \ref{fgr:sample_euler_phase}a), where a sophisticated amalgamation of fibrous and bulk grains is observed. Such hybridized configurations, though algorithmically derived, hint at the possibility of obtaining undiscovered microstructures. The breadth of microstructural diversity achievable through our generative approach is indispensable for the exploration of innovative material designs and for gaining insights into the behavior of materials subjected to varied processing environments.

To quantitatively evaluate the generated microstructure dataset, we employed multiple metrics, including Fréchet Inception Distance (FID)\cite{heusel_gans_2017}, volume fraction distributions and average two-point correlation, which provided comprehensive insights into the quality and diversity of the 50,048 generated microstructures. The FID scores for the generated dataset were approximately 40 for grain structures and 46 for phase structures\ref{tbl:generation_FID}, indicating a notable level of similarity to the statistical properties of the training dataset. Further analysis of the distribution of volume fraction\ref{fgr:sample_euler_phase}c), the generated dataset closely mirrors the training data, with both distributions centered at the same value. However, the generated dataset shows a more Gaussian-like distribution, with a slightly higher frequency of values in the 70\%-80\% range, while the training data exhibits deviations from Gaussian behavior, particularly in the 30\%-40\% range. This discrepancy may reflect the model's tendency to generate more regular and balanced structures, leading to a smoothing of irregularities present in the training dataset. In terms of the average two-point correlation\ref{fgr:sample_euler_phase}d), which quantifies spatial phase arrangements within the microstructure, revealed overall consistency between the generated and training datasets at short distances, though the generated dataset exhibited slightly higher correlation values at larger distances, suggesting the generative model may introduce subtle long-range correlations due to interpolation in latent space. 

Although the results indicate good model performance, it is necessary to discuss some of the observed discrepancies and potential limitations. Firstly, the FID scores, although generally good, may not fully reflect the model's capabilities due to two key factors: the model used to compute the FID was not specifically designed for microstructural image analysis and may overestimate the variance due to its inability to capture the nuances of the microstructural patterns; and the use of an autoencoder within the SD framework may result in the loss of some of the finer perceptual details during encoding and decoding, resulting in relatively deteriorated FID scores. However, it is worth noting that, despite these limitations, the model is still able to generate high-quality microstructures that are very similar to the training data.

The differences observed in the volume fraction distributions provide additional insights into the model's behavior. The Gaussian-like distribution of the generated dataset, along with higher frequencies in the 70\%-80\% range, reflects the model's ability to generate more balanced and regular structures. Although the training dataset exhibited non-Gaussian behavior, particularly in the 30\%-40\% range, the model’s bias toward a more regular distribution could be seen as an improvement, smoothing out the irregularities present in the training data. As the volume fraction plays a crucial role in the spatial distribution of phases, this regularization in the generated data likely contributes to differences observed in the two-point correlation.

For the two-point correlation results, the slight differences at larger distances suggest that the model may have introduced smoother or more extended correlations, possibly due to its underlying spatial interpolation, which tends to smooth out local features and introduce correlations at longer distances. While this could be seen as a bias, it is important to recognize that the model may actually be addressing potential irregularities in the training data. The training dataset itself suffers from some distributional inconsistencies,  particularly at smaller volume fractions, and the ability of the model to produce smoother correlations may indicate a more general and consistent model generation process, potentially mitigating some of these inconsistencies. Rather than seeing this as a limitation, it should be interpreted as an advantage of the model in generating microstructures with more controllable and desirable properties, thereby potentially addressing the shortcomings of the original dataset.

In conclusion, while there are some slightly discrepancies between the generated data and the training data, these differences may highlight the model’s ability to overcome certain limitations in the dataset itself. The generative process produces microstructures that are not only statistically similar to the training data but may also introduce improvements in terms of structure regularity and phase distribution, which highlights the robustness of our approach in generating realistic microstructures.

\begin{table}[h!]
\small
  \caption{\ FID scores for 50,048 randomly generated grain and phase datasets associated with the training and validation sets, respectively.}
  \label{tbl:generation_FID}
  \begin{tabular*}{\textwidth}{@{\extracolsep{\fill}}lcccc}
    \hline
    Type & Grain Train & Grain Validation & Phase Train & Phase Validation   \\     
    \hline
    FID & 40.112 & 40.243 & 46.474 & 46.553 \\
    \hline
  \end{tabular*}
\end{table}

\subsection{Advanced microstructural design with targeted material properties via ControlNet}

While SD models have the ability to produce a large number of microstructures with high fidelity, it is critical to implicitly control the model's generation of microstructures with specific properties on demand. In this section, we illustrate how control of microstructured material properties can be realized by the SD model in combination with ControlNet, using as examples both a high-precision prediction of grain orientation based on phase data and the generation of microstructures based on target coercivity.

The ControlNet is employed, augmented with a fine-tuning process on a pre-trained SD model, to generate grain orientation images from phase data and microstructures with specific coercivity values. For grain orientation prediction, the training dataset consisted of 10,000 randomly selected phase images and their corresponding grain images from the original SD model training set, enabling a targeted fine-tuning of the model. Specifically, the input to the fine-tuned model is a single-channel phase image, while the output is a composite four-channel image. To further enhance ControlNet’s capabilities, we refined the model architecture to accept not only image-based inputs but also numerical or vector-based property inputs. This modification allows ControlNet to generate microstructures with tailored physical properties, such as coercivity. The training dataset for this task included 10,000 randomly selected images from the SD model training set, along with their corresponding coercivity values calculated using MuMax3 (applied to dual-phase $SmCo_5/Sm_2Co_{17}$ composites as a hypothetical material system). As a result, the fine-tuned ControlNet model can now generate microstructures with specific coercivity ranges, offering a powerful tool for the customized design of materials.

\begin{figure}[h!]
\centering
  \includegraphics[width=16cm]{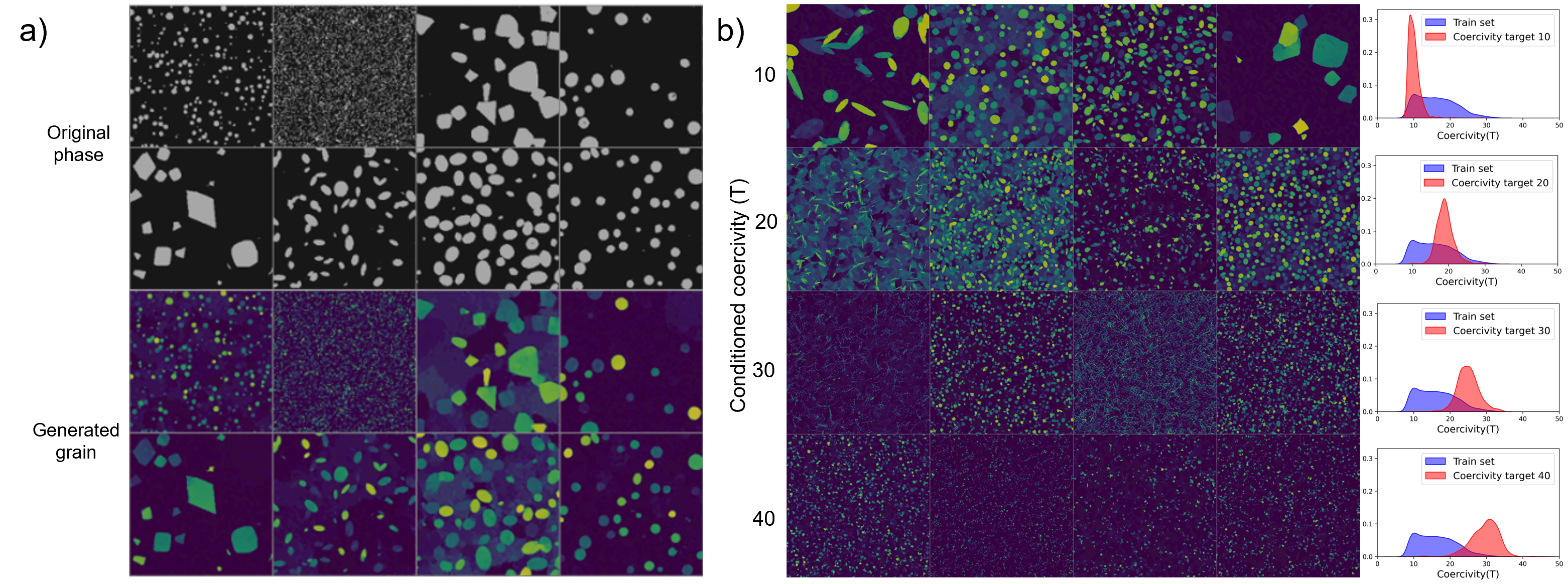}
  \caption{The a) input phase image and b) the conditioning generated grain images using ControlNet.}
  \label{fgr:Controlnet}
\end{figure}

The comparative analysis between the input phase images and their corresponding generated grain images, as shown in Figure \ref{fgr:Controlnet} a), demonstrates the exceptional consistency of our model in predicting grain orientations across various microstructural regions. The model shows remarkable accuracy in delineating distinct (on grain boundaries) and coherent (within individual grains) grain orientations at critical junctures. These regions are essential for understanding material anisotropy and inferring material properties. A quantitative evaluation of the model's performance was conducted by calculating the FID score for 27,696 data points from the validation set. The FID score comparing the original and generated grain images was 41.464, which is consistent with results typically observed in SD model generation tasks. This indicates a reasonable level of similarity, considering the complexity of the task. The model’s proficiency in capturing and replicating complex orientation gradients and textures, which are fundamental to the material's structure, provides valuable insights into the material's overall properties and behavior.

The results of generating microstructures conditioned on coercivity values of 10T, 20T, 30T, and 40T are shown in the left column of Figure \ref{fgr:Controlnet} b), from top to bottom. For each condition, 10,000 microstructures were generated, and their coercivities were calculated using MuMax3. As the conditioned coercivity increases, it can be observed that the proportion of the $SmCo_5$ phase increases, while the $Sm_2Co_{17}$ grain size becomes smaller and more uniformly distributed. This is consistent with the physical characteristics of the two phases: $SmCo_5$, with its higher uniaxial anisotropy constant, contributes to higher coercivity, while the smaller and more uniform $Sm_2Co_{17}$ grains enhance the overall stability of the microstructure by increasing the pinning of domain walls. The increase in $SmCo_5$ phase content and the reduced $Sm_2Co_{17}$ grain size together result in an optimized microstructure for resisting magnetic reversal, which corresponds to the observed increase in coercivity. The right column of the figure shows the comparison of coercivity distributions between the generated microstructures (in red) and the ControlNet model's training set (in blue). As the conditioned coercivity increases, the red distribution of the generated results shifts positively along the X-axis, indicating an overall increase in coercivity. However, this shift becomes more gradual as the conditioned coercivity approaches the upper boundary of the original dataset. This moderation likely occurs because the training set contains fewer high-coercivity samples, limiting the model's ability to generate highly coercive microstructures in those regions. Despite this limitation, the model successfully produced microstructures with coercivities exceeding the maximum values found in the training data, confirming that the generative model, when combined with ControlNet, has the capacity to extrapolate beyond the boundaries of the training data, discovering new microstructures with superior magnetic properties.

In conclusion, microstructure generation methods combining ControlNet with Stable Diffusion models show significant advantages, particularly in generating microstructures with specific physical properties and accurately predicting grain orientation. Compared to traditional statistical methods such as Voronoi splitting and Monte Carlo simulation, our approach has advantages in terms of flexibility, automation capability and generation performance. Traditional methods usually rely on simple geometric segmentation or probabilistic models, which struggle to capture the complex curvature of grain boundaries and the orientation gradients within grains, and are often accompanied by tedious manual operations and prone to subjectivity. In contrast, our method leverages advanced machine learning algorithms trained on a comprehensive dataset of phase images and physical properties, allowing the model to autonomously learn and generate microstructures with intricate orientation variations and anisotropic features. Moreover, by conditioning the model on specific physical properties, we can precisely control material characteristics and extend beyond the limits of the training data to discover new material structures that are challenging for conventional methods to capture.

\section{Conclusions}

In this paper, we have harnessed the capabilities of SD models and ControlNet to address complex challenges in microstructure analysis, spanning microstructure reconstruction, interpolation, inpainting, generation, and the precise reconstruction of grain orientations from phase information. Our findings illuminate the significant potential of these models as instrumental tools in the realm of materials science, offering nuanced insights into the intricate structures of materials and their properties.

The adept reconstruction of microstructures by our model underscores its exceptional ability to replicate detailed material structures, thereby affirming its utility in advancing materials science research. The process of interpolating between microstructures has revealed the model's capacity to simulate the dynamic transformations and transitions inherent in material properties, thereby providing a novel pathway for the design of materials with tailored intermediate characteristics. Moreover, the model's proficiency in microstructure inpainting has been demonstrated, highlighting its utility in rectifying incomplete or imperfect microstructural data, a common challenge in material characterization. This capability is particularly noteworthy for its application in real-world scenarios where data may be compromised or incomplete. The generation of novel microstructures through our model, leveraging a vast synthetic dataset, opens new frontiers in the exploration of material configurations. This aspect of our research points towards the possibility of discovering materials with novel properties, thus expanding the boundaries of current material science knowledge and application. Central to our study is the implementation of ControlNet in SD model for grain orientation reconstruction based on phase information, and the microsturcture generation conditioned on desired coercivity. Compared to traditional reconstruction techniques, this method significantly reduces the human effort required in microstructure design while enhancing the control over material properties and fostering innovation.

Collectively, these contributions highlight the transformative potential of integrating advanced SD models in the field of microstructure design. By bridging the gap between ML and practical applications, our approach paves the way for the design of materials with optimized properties, thereby fostering innovation and broadening the scope of materials science research and its applications.

\section*{Conflicts of interest}
There are no conflicts to declare.

\section*{Author contributions}
Yixuan Zhang: methodology, data collection, coding, writing – original draft; Teng Long: conceptualization, writing – review and editing; Hongbin Zhang: conceptualization, funding acquisition, writing – review and editing, supervision.

\section*{Data and code availability}
All data needed to produce the work are available upon reasonable request from the corresponding author. All codes generated or used during the study are available in the github repository. All model weights generated during the study are available in the huggingface repository.

\section*{Acknowledgements}
The authors would like to express their sincere gratitude to Manuel Richter for providing the SOC strength data and for his valuable discussions.
This work was also supported by the Deutsche Forschungsgemeinschaft - Project-ID~405553726 - TRR 270.
The authors gratefully acknowledge the computing time provided to them on the high-performance computer Lichtenberg at the NHR Centers NHR4CES at TU Darmstadt. This is funded by the Federal Ministry of Education and Research, and the state governments participating on the basis of the resolutions of the GWK for national high-performance computing at universities (https://www.nhr-verein.de/unsere-partner).

%%%%%%%%%%%%%%%%%%%%%%%%%%%%%%%%%%%%%%%%%%%%%%%%%%%%%%%%%%%%%%%%%%%%%
%% The appropriate \bibliography command should be placed here.
%% Notice that the class file automatically sets \bibliographystyle
%% and also names the section correctly.
%%%%%%%%%%%%%%%%%%%%%%%%%%%%%%%%%%%%%%%%%%%%%%%%%%%%%%%%%%%%%%%%%%%%%
\bibliography{citation}

\end{document}